\documentclass[useAMS,usenatbib]{mn2e}
\usepackage{graphicx,amsmath,multirow,amssymb}
\usepackage{natbib}
\usepackage{epstopdf}
\newcommand\apj{{ApJ}}%
\newcommand\apjl{{ApJ}}%
\newcommand\apjs{{ApJS}}%
\newcommand\aap{{A\&A}}%
\newcommand\mnras{{MNRAS}}%
\newcommand{\comment}[1]{}
\newcommand\ocen{$\omega$~Cen~}

\newcommand\Msun{M$_\odot$}
\def\simgt{\lower.5ex\hbox{$\; \buildrel > \over \sim \;$}}
\def\simlt{\lower.5ex\hbox{$\; \buildrel < \over \sim \;$}}

\title[HBB in super-AGB stars]
{Hot Bottom Burning in the envelope of SAGB stars}
\author[P. Ventura, F. D'Antona]{P. Ventura$^{1}$\thanks{E-mail:
ventura@oa-roma.inaf.it (AVR)} and F. D'Antona$^{1}$\\
$^{1}$INAF-Osservatorio Astronomico di Roma, Via Frascati 33, Monte Porzio Catone 00040,
Italy}
\begin{document}

\date{Accepted . Received ; in original form }

\pagerange{\pageref{firstpage}--\pageref{lastpage}} \pubyear{2002}

\maketitle

\label{firstpage}

\begin{abstract}
We investigate the physical and chemical evolution of population II stars 
with initial masses
in the range 6.5-8 M$_{\odot}$, which undergo an off centre carbon ignition under
partially degenerate conditions, followed by a series of thermal pulses,
and supported energetically by a CNO burning shell, above a O-Ne degenerate core.
In agreement with the results by other research groups, we find that the O-Ne
core is formed via the formation of a convective flame that proceeds to the
centre of the star. The evolution which follows is strongly determined by the
description of the mass loss mechanism. Use of the traditional formalism with the
super-wind phase favours a long evolution with many thermal pulses, and the
achievement of an advanced nucleosynthesis, due the large temperatures reached
by the bottom of the external mantle. Use of a mass loss recipe with a strong
dependence on the luminosity favours an early consumption of the stellar envelope, 
so that the extent of the nucleosynthesis, and thus the chemical composition of
the ejecta, is less extreme. The implications for the multiple populations in
globular clusters are discussed.
If the ``extreme'' populations present in the most massive clusters are a
result of direct formation from the super-AGB ejecta, their abundances may constitute
a powerful way of calibrating the mass loss rate of this phase. This calibration 
will also provide informations on the fraction of super-AGBs exploding as single
e-capture supernova, leaving a neutron star remnant in the cluster.

\end{abstract}


\section{Introduction}
The deep spectroscopic surveys of Globular Clusters (GC) stars performed
in the last decades have revealed star-to-star differences, that trace
well defined abundance patterns, involving all the ``light'' elements,
up to Aluminium \citep{norris}. The discovery that stars belonging to the main
sequence or the sub--giant branch show the same patterns 
of abundances present in the red giants of the same cluster \citep{carretta1}, 
indicated that some self-enrichment mechanism has been active in GCs, and 
that a new, second, stellar generation (SG) formed, from the ashes of the evolution of 
older objects, belonging to a first generation (FG). 
A key-point towards the understanding of the processes behind these 
observational results is the large fraction of stars with the anomalous
chemistry, that hardly drops below 50\% of the total number of
stars examined, and exceeds 80\% in some clusters \citep{carretta2, carretta3} .
The Oxygen-Sodium anticorrelation is by far the most studied, and
has been confirmed to exist in practically all the GCs investigated,
although the slope of the O-Na relationship, and the lowest oxygen 
abundances detected, vary considerably from cluster to cluster.
In many clusters a Mg-Al anticorrelation
is also detected, although the extent of the magnesium depletion is
still debated \citep{cohen}. Unlike the O-Na trend, the Mg-Al relationship
is not of straight interpretation, because not only the slope (if any),
but also the maximum magnesium and the minimum aluminium mass fractions, 
which are commonly interpreted as the abundances of the primordial population, 
change from cluster to cluster \citep{carretta3}.
Photometric signatures are generally less striking than the spectroscopic evidence, apart from
the presence of very anomalous "blue" main sequences (bMS) in two among the most massive
clusters \citep[\ocen\ and NGC~2808][]{bedin2004,piotto2005,dantona2005,piotto2007} and
by the presence of very extended horizontal branches and very luminous RR~Lyrae in two
metal rich clusters --NGC 6388 and NGC 6441. Both these features indicate the presence 
of a very helium rich population (Y$\simgt$0.36--0.38) in these four clusters \citep[e.g.][]{norris2004, caloi2007}. 

Three distinct scenarios have been proposed so far for the progenitors
of the SG, each of which must face some difficulties in reconciling the 
theoretical predictions with the observational evidence: a) winds from massive 
AGBs \citep{paolo01}, b) ejecta from fastly rotating massive stars 
\citep{decressin}, and c) massive binaries \citep{langer}. All these
pollutors together, plus the possible contribution of stellar collisions, have been
proposed by \cite{sills}.
In this work we focus on the first hypothesis, already
outlined on qualitative grounds by \citet{paolo08} and \citet{paolo09}, 
and fixed on a most robust framework by two seminal papers, computing  first
the possible hydrodynamical and N--body evolution,
and further exploring the chemical evolution of protoclusters \citep{annibale,annibale2010}. 
In these latter works, it is suggested that the SG forms in a cooling flow established in the GC core by the gas ejected at low velocity by massive AGBs. 
This process is restricted to a narrow time interval $\simlt 10^8$yr, thus allowing only the contribution of stars with
masses M$\geq 5$M$_{\odot}$, after which star formations stops, due to the onset
of SNIa explosions. The loss of stars from the outskirts of the
cluster helps diminishing the FG/SG stars number ratio, because this mechanism
hardly involves, in a first phase, the core born SG objects.

\citet{annibale}, following \cite{pumo2008}, also suggested that 
in the most massive clusters a preminent role 
in the star forming process is played by super-AGB (SAGB) stars, 
defined as objects with initial masses in the range 
(9-11M$_{\odot}$)\footnote{The range of masses involved depends on the assumptions
regarding the overshooting from the convective core during the two
main phases of nuclear burning: overshooting increases the size
of the convective core, and thus shifts the masses involved to
lower values.} that undergo off--center carbon ignition in partially 
degenerate conditions, and end-up their evolution as O-Ne white dwarfs. 
They showed, by means of hydrodynamical simulations, that stars belonging to the very high
helium populations quoted above, found only in
 the most massive clusters, are possibly formed {\it directly} from
the winds of the SAGBs, before mixing with pristine gas favours a less extreme
chemistry. SAGBs, according to \cite{siess07}, have 
helium envelope abundance Y=0.36 -- 0.38 after the second dredge up, and as such 
are the only viable candidate for the bMSs. Unlike helium, the other yields of 
SAGBs are the result of mass loss and hot bottom burning (HBB) in the phase of 
thermal pulses following the formation of the O--Ne core. Common interpretation 
attributes to them the most extreme chemical anomalies found in the clusters 
having the bMS \citep[see, e.g.][]{annibale2010}, but a full confirmation will 
have to wait for accurate abundance determinations of stars belonging to the bMSs.
It is well known that the massive AGB yields are dramatically dependent of several 
uncertain input physics parameters, such as convection model, mixing and nuclear 
cross sections \citep{paolo05a, paolo05b}, and we can expect similar, or even more 
extreme, uncertainties in the SAGB mass range.

Theoretical models of SAGBs have been first presented in a series of paper
by \citet{garcia1}, \citet{iben}, \citet{ritossa1}, \citet{ritossa2}, 
until the recent updates by \citet{siess06} and \citet{siess07}: the carbon 
burning phase in these models is described in terms of a burning flame, 
that propogates inwards, until carbon burning reaches the center. 
The main effects due to overshooting
\citep{gil}, and thermohaline mixing \citep{siess09} were also
investigated and discussed. The recent work by \citet{siess10} presents
the first database of yields of SAGBs with different mass and metallicity,
and include also the Thermal Pulses (TP) phase that follows the
extinction of carbon burning.
Finally, the investigation by \citet{doherty} outlined the
robustness of the theoretical description of these evolutionary phases,
showing that when a sufficiently accurate temporal resolution is adopted,
results obtained with different codes show a satisfactory agreement.

\cite{siess10} SAGB computations are based on physical assumptions very different from our own \citep{paolo09}, that we have already applied to the interpretation of GC abundance patterns. Here we extend our models to the range of SAGB masses, to show how the results depend on detailed physical inputs and, mostly, on the assumed mass loss law. In fact,
on the chemical side, we expect uncertainties on the yields' computation,
because the mass loss description, the efficiency of the
convective modelling, the choice of the nuclear cross-sections, are all
expected to play a role in the physical, and thus chemical description
of the evolutionary phases following the formation of the O-Ne core.
The fate of these stars is also highly uncertain, as it depends on 
the velocity with which the convective mantle is consumed, and, in
more details, whether this process is completed before electron
captures begin inside the core \citep[e.g.][]{poelarendes}.

Our aim is thus to explore the uncertainties
associated to the yields of these stars, and particularly to the
oxygen and sodium mass fractions, because mostly these two elements are
investigated in the spectroscopic surveys of GC stars.
We will also consider the helium content in the ejecta, and 
the magnesium and aluminum yields, following the scheme given in \cite{paolo09}
\citep[but see also][]{annibale2010}.

\section{Hot Bottom Burning: the different fate of Oxygen, Sodium
Magnesium and Aluminium}
Massive AGBs do not obey to the classic luminosity vs.
core mass relationship found by \citet{paczynski}, because this latter
is based on the assumption that a radiative buffer is present between
the H-burning shell and the bottom of the convective envelope, 
whereas in massive AGBs the bottom of the convective envelope
eventually overlaps with the H-burning shell, so that part of
the nuclear energy is generated directly into the external mantle
\citep{blo2}.
This phenomenon, normally indicated as Hot Bottom Burning, is
of paramount importance for the topic of the self-enrichment by
massive AGBs, because it is an efficient way of polluting the
stellar environment with matter that was nuclearly processed in
repeated proton-capture reactions \citep{cottrell,paolo01}.

In a series of papers, \cite{paolo05a, paolo08, 
paolo09} showed that HBB is expected to operate efficiently in all 
popII AGB models with initial masses M $\simgt$ 4M$_{\odot}$, provided that 
a high efficiency convection model is adopted, e.g. the Full
Spectrum of Turbulence description for turbulent 
convection  \citep{canuto}. The surface chemistry is consequently modified.

Oxygen is burnt by proton capture; the rate of destruction increases
for higher temperatures, and is thus stronger in the more massive
models, whose core mass is larger. In the most massive AGBs, i.e. the
highest masses not undergoing carbon ignition \citep[$\sim 6-6.3M_{\odot}$ in][] 
{paolo08}, the final oxygen abundance is a factor of $\sim 20$ 
smaller than the initial mass fraction \citep[see Figure 2 in ][]{paolo08}, 
and the average oxygen in the ejecta is decreased by a factor $\sim 7$. 
When M increases, the trend with mass tends to flatten, and eventually
is reversed \citep[see Table 2 in ][]{paolo09}, because the 
higher temperatures, that tend to decrease the oxygen content of the 
ejecta, are partly compensated by the larger mass loss, suffered by the 
most massive models in the early AGB phases, when the surface oxygen 
was still rather large. This is an effect of the steep increase of mass 
loss with luminosity of the \citet{blo} recipe.

The behaviour of sodium is more complex, as it is a result of 
a production channel and two destruction reactions, with
very different sensitivity to the temperature in the range 
of interest here \citep{hale1,hale2}. For $T < 60$MK the reaction of 
proton capture by $^{22}$Ne nuclei dominates, thus producing great 
amounts of sodium; at higher temperatures, the destruction channels via proton capture
take over, and sodium is destroyed, reaching an equilibrium value,
at which creation and burning rates cancel each other
(see Fig.4 in \citet{paolo08}). Two consequences arise from the 
above description:

\begin{itemize}

\item{
Unlike oxygen, the predictions concerning the sodium yields are 
not robust, given the uncertainties of the p-capture reactions of the 
Ne-Na cycle.
}

\item{
The temperature sensitivity of the $^{23}$Na+p reaction, much steeper compared
to the production channel, if confirmed, is necessarily associated to a direct 
correlation between the average oxygen and sodium content of the ejecta  
\citep{pavel,herwig04,karakas2007,paolo08}; only in the very early AGB phases, 
when oxygen burning is active and the sodium production prevails over destruction, 
oxygen and sodium show an opposite behaviour.
}

\end{itemize}

HBB modifies the surface abundances of magnesium and aluminium too.
$^{24}$Mg is destroyed when T$_{bce} \sim 60-70$MK, and $^{27}$Al is
produced by a proton-capture chain, that also increases the
surface mass fractions of the two heavy magnesium isotopes, $^{25}$Mg
and $^{26}$Mg. \citet{paolo08} find that the extent of the Al production
(hence, of the depletion of magnesium) increases with mass in massive AGBs,
though an upper limit of [Al/Fe] $\sim 1$ is reached, which is again due
to the high mass loss experienced by the most massive stars at the
beginning of the AGB phase. These results were obtained by adopting the upper
limits for the two proton capture reactions by the heavy magnesium isotopes,
although the recent analysis by \citet{annibale2010} on M4 recommends the
use of the standard cross sections. 
The uncertainties associated to the relevant cross-sections of the Mg-Al 
chain render the results uncertain by $\sim 0.3$ dex (\citet{paolo08}, 
but see also the discussion in \citet{izzard}).

\section{The models}
\subsection{Physical and chemical inputs}
To investigate the nucleosynthesis at the bottom of
the external mantle of SAGB stars after the carbon burning
phase, we calculate models with the standard chemistry 
typical of intermediate metallicity GCs, i.e. Z=0.001,
Y=0.24, and an $\alpha$-enhanced mixture, with $[\alpha/$Fe]=+0.4
\citep{gs98}.

The nuclear network includes the most important reactions
involving all the chemical species up to silicon. Most of the 
cross-sections were taken from the NACRE compilation \citep{angulo};
to maintain consistency with \citet{paolo05a, paolo05b, paolo08}
we used the upper limits for the proton capture reactions by
$^{25}$Mg and $^{26}$Mg. This choice maximizes the aluminum production;
a full description of the Mg-Al production and destruction will be given in
a forthcoming paper. The reaction rates of the $^{14}$N(p,$\gamma$)$^{15}$O
reaction were taken from \citet{formicola}, whereas the cross sections
by \citet{hale1} and \citet{hale2} were adopted for the reactions of
the Ne-Na cycle. Sodium production was maximized by adopting the upper
limit for the proton capture reaction by $^{22}$Ne nuclei.

Mass loss was modelled according to \citet{blo}, while
some comparison sequences follow the recipe by \citet{VW93}. 
Nuclear burning and convective
mixing were coupled by means of a diffusive approach 
\citep{clout}.
Convective eddies are allowed to overshoot into regions of
radiative stability during both hydrogen and helium nuclear
burning in the core; this extra-mixing is simulated by an exponential
decay of velocities beyond the formal border fixed via the
classic Schwartzschild criterium, with an e-folding distance
$\zeta H_p$, with $\zeta=0.02$, in agreement with a calibration
based on the observed widths of the main sequences of open
clusters, presented in \citet{paolo1}.
The range of masses investigated is 6.5-8 solar masses; this range 
is to be compared with the interval 8-9M$_{\odot}$ found for the
same chemistry by \citet{siess10}.
The different assumptions on overshooting are the cause of
the difference in the mass range of SAGBs.
The involved core masses in fact are very similar
in the two investigations (see Figure  \ref{f1}). 
The results are also in agreement with the exploration on the 
effects of overshooting by \citet{gil}. 

\subsection{The Carbon burning phase}
\begin{table*}
\caption{Carbon burning properties}
\label{sagbfis}
\begin{tabular}{cccccc}
\hline
M/M$_{\odot}$ & $\tau_{CB-HeB}(10^5)$yr & M$_{CO}$ & m$_{ign}$ & L$_C$ & M$_{SDU}$ \\
\hline
6.5 & 2.3 & 1.06 & 0.60 & 7.1(6) & 1.15 \\
7.0 & 2.0 & 1.12 & 0.50 & 2.3(6) & 1.20 \\
7.5 & 1.7 & 1.18 & 0.36 & 1.0(6) & 1.30 \\
8.0 & 1.4 & 1.29 & 0.21 & 6.3(5) & 1.40 \\
\hline
\end{tabular}
\end{table*}
The main physical properties concerning the C-burning phase are 
presented in Tab.~\ref{sagbfis}, where for the various masses involved
we show the time elapsed from the end of the core-He burning phase to the
ignition of the first C-burning episode, the mass of the CO core when
carbon burning starts, the mass of the layer at which carbon is ignited,
the maximum luminosity produced, and the innermost layer reached by the 
bottom of the surface convective zone during the second dredge-up.

\begin{figure}
\begin{minipage}{0.47\textwidth}
\resizebox{1.\hsize}{!}{\includegraphics{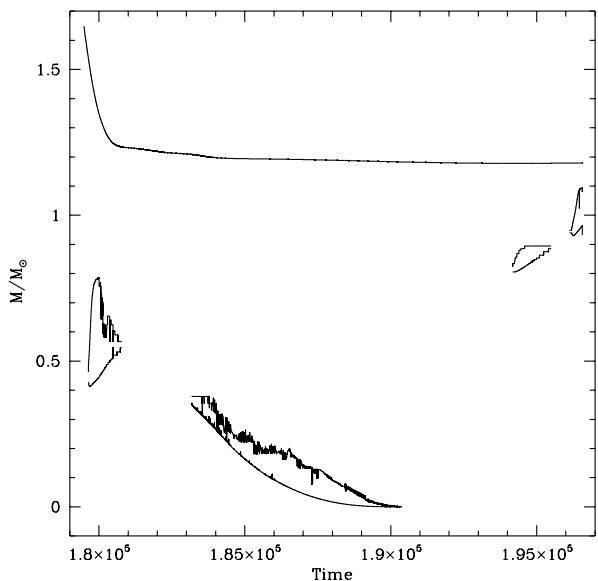}}
\end{minipage}
\caption{Kippenhan diagram showing the temporal behaviour of the
development of convective regions during the C-burning phase of
a 7M$_{\odot}$ model.
}
\label{kipp}
\end{figure}

\begin{figure}
\begin{minipage}{0.47\textwidth}
\resizebox{1.\hsize}{!}{\includegraphics{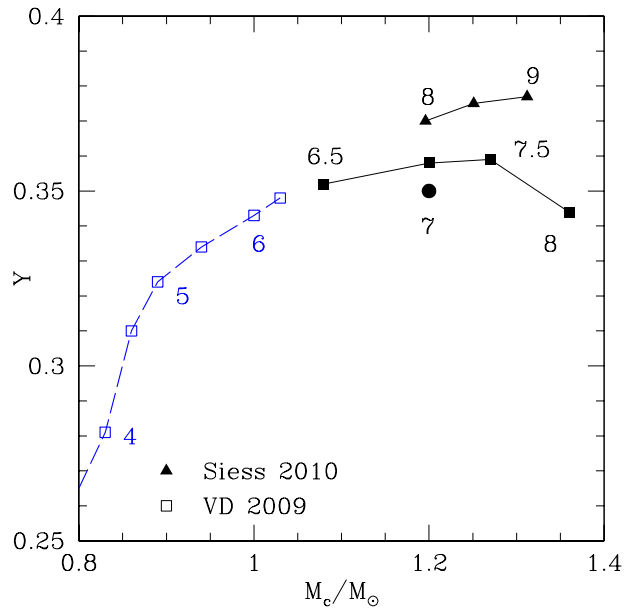}}
\end{minipage}
\caption{The helium content of the
ejecta of the massive AGB \citep[open squares, from ][]{paolo09} 
and of the SAGB models calculated in this work  
(full squares) is shown as a function of the core mass. The numbers
next to the points indicate the initial mass of the model.
The values for the SAGBs by \citet{siess10} (full triangles) for similar chemistry are shown.
The full dot indicates how the helium content from the 7M$_{\odot}$ decreases when a shallower mass loss formulation is adopted. 
}
\label{f1}
\end{figure}

All the models undergo an off-center carbon ignition; the point at which 
the nuclear energy release is highest is more internal the higher is the 
mass, ranging from $\sim 0.6M_{\odot}$ for M=6.5M$_{\odot}$, down to 
$\sim 0.2M_{\odot}$ for M=8M$_{\odot}$. 
This early phase of carbon burning is followed by the development of a 
convective flame, that proceeds inwards. In all the models investigated, 
this instability region eventually reaches the center, and forms a 
O-Ne core, with the only exception of the 6.5M$_{\odot}$ model, that is left 
with a core made up of carbon and oxygen;  in this latter model carbon burning
is aborted, in analogy to the behaviour of the masses just above the limit for carbon
ignition, described by \citet{doherty}. The formation of the O-Ne core is 
followed by other off-center carbon burning episodes, whose intensity becomes 
progressively lower, after which the TPs phase begins. Fig.~\ref{kipp} shows
a Kippenhahn diagram for the model with mass 7M$_{\odot}$, indicating
the borders of the convective shells that form as a consequence of carbon
ignition, and the base of the external convective mantle. A detailed analysis
of the C-burning phase is beyond the scopes of the present work: the interested 
reader may find in \citet{siess06} a detailed physical description of the 
C-burning phase in super-AGB stars.

\begin{table*}
\caption{Properties of SAGB models and average abundances in the ejecta}
\label{sagbmod}
\begin{tabular}{ccccccccccccc}
\hline
\hline
M/M$_{\odot}$ &  M$_c/$M$_{\odot}$  &  $T_{\rm bce}^{\rm max}/10^8$ & Y & 
$[^{16}$O/Fe] & [Na/Fe] & [Mg/Fe] & $[^{27}$Al/Fe] & A(Li) &  $\dot{M}$ law &
M$_{\rm tot}^f/$M$_{\odot}$ & M$_c^f/$M$_{\odot}$ & NTP \\
\hline
6.5 &  1.08 & 1.16 & 0.352 & -0.24 & 0.32 & 0.226 & 0.80 & 2.36 &B95 & 1.70 & 1.10 & 32 \\
7.0 &  1.20 & 1.20 & 0.358 & -0.15 & 0.39 & 0.253 & 0.74 & 2.12 &B95 & 1.99 & 1.23 & 38 \\
7.5 &  1.27 & 1.27 & 0.359 &  0.01 & 0.67 & 0.289 & 0.57 & 2.75 &B95 & 2.15 & 1.29 & 48 \\
8.0 &  1.36 & 1.47 & 0.344 &  0.20 & 1.00 & 0.290 & 0.40 & 4.39 &B95 & 2.20 & 1.38 & 53 \\
\hline
7.0 &  1.20 & 1.34 & 0.350& -0.69 & 0.17 & 0.09  & 0.54 & 2.58 &VW93 & 2.15 & 1.24 & 348\\
\hline
 \end{tabular}
\end{table*}

\subsection{How much helium is produced by SAGBs?}
The second dredge-up is a common feature of all the models investigated;
the evolutionary stage at which it takes place depends on the initial mass: it
is prior to C-burning in all the models, apart form the 8M$_{\odot}$ case,
in which it occurs after the formation of the ONe core. This behaviour
is in agreement with the investigation by \citet{garcia2}. On the chemical side,
the main consequence of the second dredge-up is the increase in the helium
mass fraction, up to $Y \simeq 0.36$ (see column 4 in Table \ref{sagbmod}). 
This is the maximum enrichment of helium that is produced by SAGBs according to the 
present investigation. It is in resonable agreement, though a bit lower, 
than the helium yields by \citet{siess10} for the same metallicity. 
Figure \ref{f1} shows the abundances
in the ejecta of the massive AGBs from \cite{paolo09} (open squares) and of the present models
(full squares), as a function of the core mass (CO cores from 4 to 6.5\Msun, ONe
cores for larger mass).

Our SAGB models do not provide helium abundances as large as Y$\sim$0.40 
invoked to interpret the bMS in \ocen \citep{sollima} and NGC 2808 \citep{piotto2007}. 
Notice however that the values depend on the interpretation of the MS colors, 
on which uncertainties may be large \citep{portinari2010}.

\begin{figure*}
\begin{minipage}{0.47\textwidth}
\resizebox{1.\hsize}{!}{\includegraphics{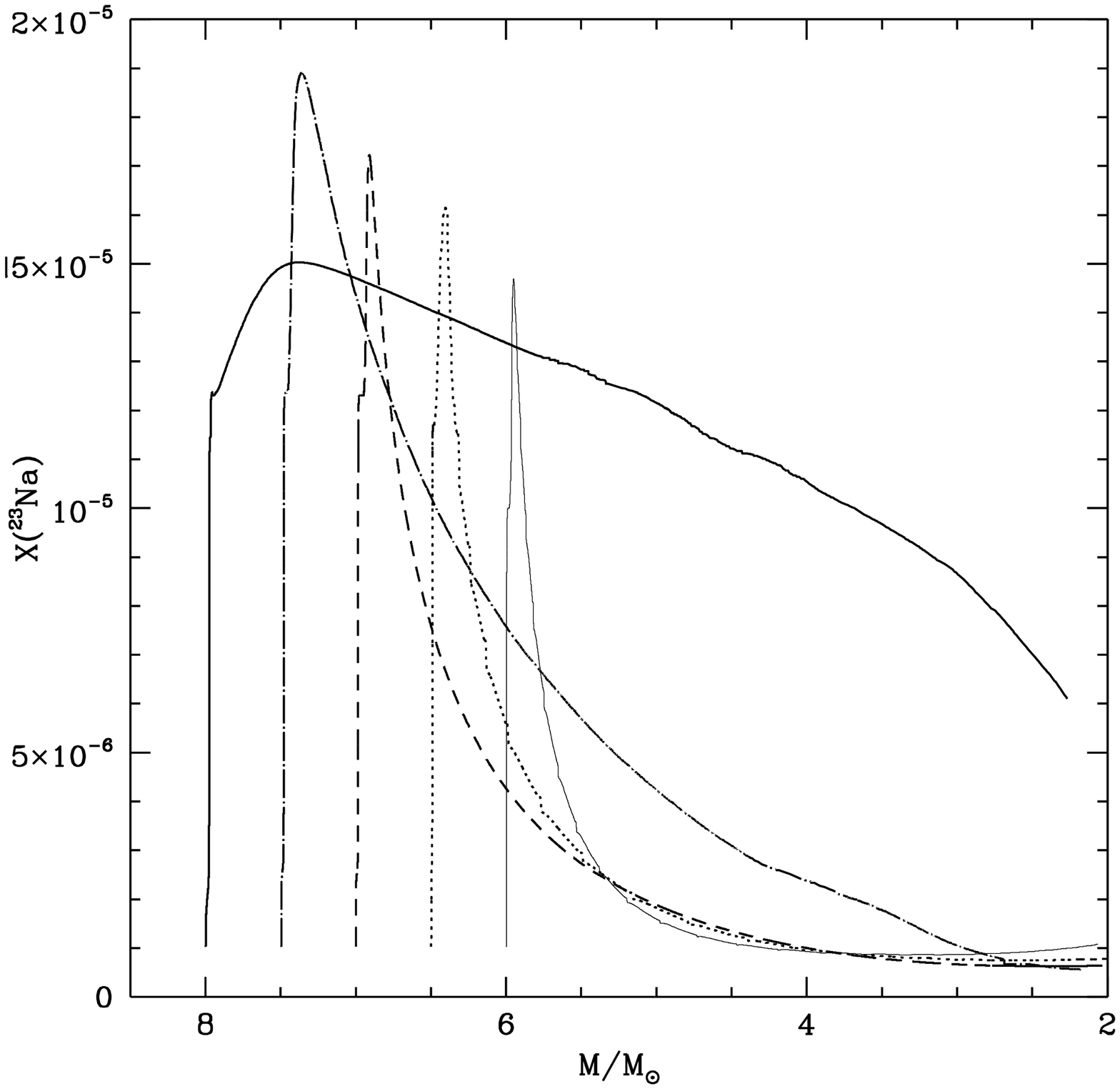}}
\end{minipage}
\begin{minipage}{0.47\textwidth}
\resizebox{1.\hsize}{!}{\includegraphics{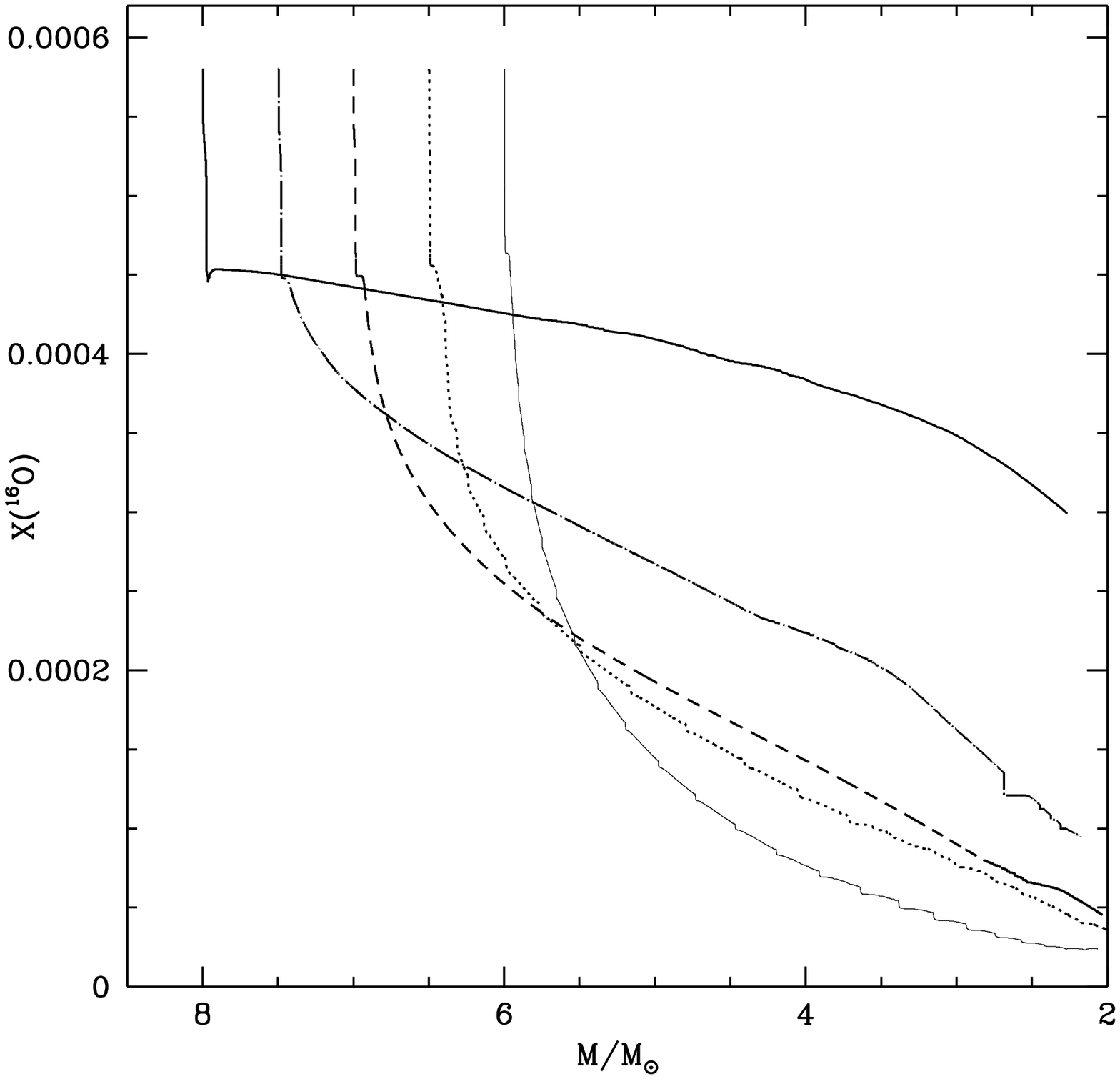}}
\end{minipage}
\caption{Left panel:  The variation during the evolution
of the sodium abundance at the surface of models with initial
masses 6M$_{\odot}$ (light solid line), 6.5M$_{\odot}$ (dotted),
7M$_{\odot}$ (dashed), 7.5M$_{\odot}$ (dot-dashed), 
8M$_{\odot}$ (solid); Right panel: the same as the left panel,
but for the oxygen surface mass fraction.
}
\label{ona}
\end{figure*}

\subsection{The effects of HBB}
Tab.~\ref{sagbmod} shows the main properties of the TP phase of
our models, i.e. the core mass at the first thermal pulse, the
maximum temperature reached at the base of the convective mantle,
the average chemistry of the ejecta, the mass of the core and of 
the whole star when the computations stopped, and the number of
thermal pulses experienced.

During the TP phase no extra-mixing was considered from any convective 
border, and no appreciable third dredge-up was experienced by any of our 
models: the surface chemistry changes only for the effects of HBB.

Fig.~\ref{ona} shows the behaviour of the surface oxygen (right)
and sodium (left), as the envelope mass is consumed. To show
the continuity with the previous investigations, limited to
models not undergoing any carbon burning, we also show the
evolution of a 6M$_{\odot}$ model, taken from \citet{paolo08}. 
Oxygen is destroyed in all models during the thermal pulses phase. 
Despite the higher temperatures attained at the bottom of the 
convective envelope in the more massive models (see Table \ref{sagbmod}), 
the higher mass loss experienced (due to the larger luminosity) favours 
a flatter oxygen vs. mass relationship, because a lot of mass is lost 
before the oxygen is burnt in great quantities. 

The average oxygen content of the ejecta, which in Table \ref{sagbmod} is 
expressed in terms of [O/Fe]\footnote{Here we use the standard notation, 
by indicating with [X/Fe] the quantity $\log(X/Fe)-\log(X/Fe)_{\odot}$.} is
found to increase with mass; this trend agrees with the previous
findings by \citet{paolo09} (most massive AGBs) and by \citet{siess10} 
(SAGBs), and indicates that the stars whose ejecta show the most extreme 
chemistries are those close to the limit for carbon ignition.

The behaviour of sodium is more tricky. The maximum sodium
produced increases with mass (as a consequence of the larger
temperatures attained at the bottom of the convective envelope), 
with the only exception of the 8M$_{\odot}$ model, that 
achieves the second dredge-up after the C-burning 
phase, and evolves rapidly to high temperatures: in this case 
sodium begins to be
destroyed before the early increase, normally associated to
the conversion of the $^{22}$Ne transported to the surface
by the second dredge-up. Similarly to oxygen, the sodium 
content of the ejecta increases with mass (see Table \ref{sagbmod}).

\begin{figure}
\begin{minipage}{0.47\textwidth}
\resizebox{1.\hsize}{!}{\includegraphics{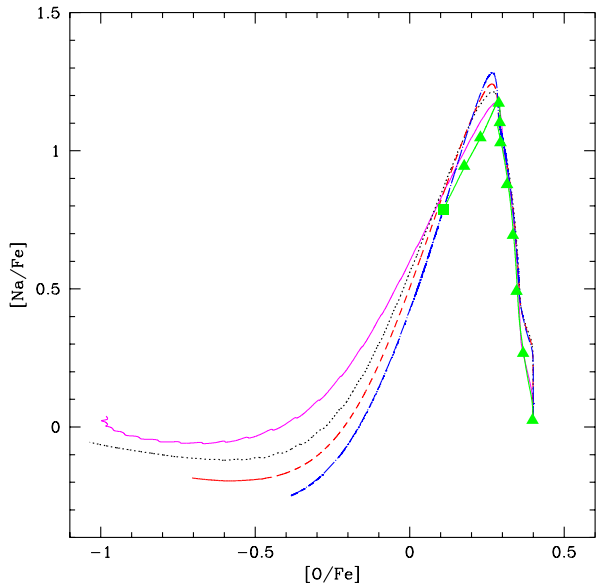}}
\end{minipage}
\caption{The simultaneous variation of the surface abundances
of oxygen and sodium in the SAGB models presented in this
investigation. The labels are the same as in the two panels
of Fig.\ref{ona}. For clarity reasons, the path followed by
the 8M$_{\odot}$ model is indicated by solid triangles, and is
terminated by the full square
}
\label{ofenafe}
\end{figure}

The simultaneous behaviour of oxygen and sodium abundance
at the surface of these stars along their evolution is shown
in Fig.\ref{ofenafe}. We used the quantity [X/Fe] on both
axis. We see that all the models follow the same qualitative
behaviour, with an initial anticorrelated phase, during
which oxygen is reduced whereas sodium increases, and a 
following phase, when both elements are destroyed at the
bottom of the external envelope. 

Oxygen is destroyed with continuity until the envelope
is removed completely, whereas sodium follows an approximately 
asymptotic behaviour, that corresponds to the afore mentioned 
balance between the production and
the destruction channels. The curves corresponding to the more
massive models (see in particular the evolution of the 8M$_{\odot}$
model, indicated by solid triangles and by the full square
pointing the end of the evolution), stop at a stage when the
oxygen and sodium abundances are still large, because the
strong mass loss favoured a rapid consumption of the whole
envelope, before the surface mass fractions of both elements 
could diminish to nominal values.

\begin{figure}
\begin{minipage}{0.47\textwidth}
\resizebox{1.\hsize}{!}{\includegraphics{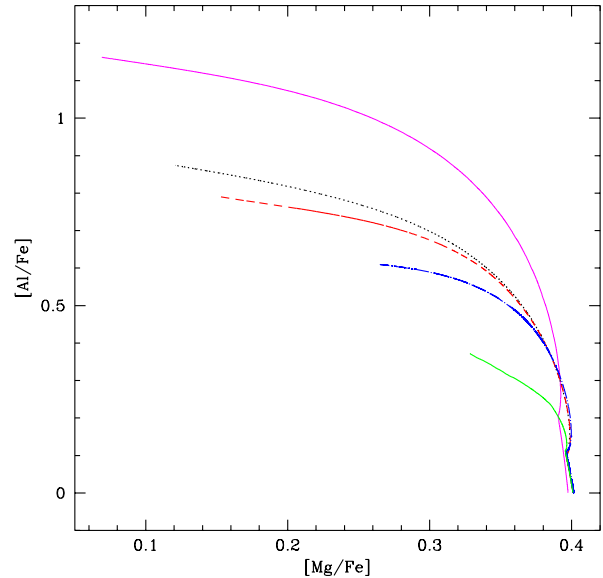}}
\end{minipage}
\caption{The simultaneous variation of the surface abundances
of magnesium and aluminium in the SAGB models presented in this
investigation. The labels are the same as in the two panels
of Fig.\ref{ona}.
}
\label{mgal}
\end{figure}

The variation of the surface contents of magnesium and $^{27}$Al
is shown in Fig.\ref{mgal}. We note that a stronger production of
aluminium is achieved in the models of smaller mass, because
in the most massive stars the strong mass loss prevents an advanced
Mg-Al nucleosynthesis, and only a modest increase in the 
surface aluminium is reached.

\section{How sensitive are the yields to the details of SAGB modelling?}
Before turning to the impact of these results on the interpretation
of the chemical anomalies observed in GC stars, and on the 
self-enrichment scenario by massive AGBs, we try to understand
how robust these results are, and their sensitivity to the
various uncertainties (convection, mass loss, cross-sections)
affecting the SAGB modelling.

\begin{figure}
\begin{minipage}{0.47\textwidth}
\resizebox{1.\hsize}{!}{\includegraphics{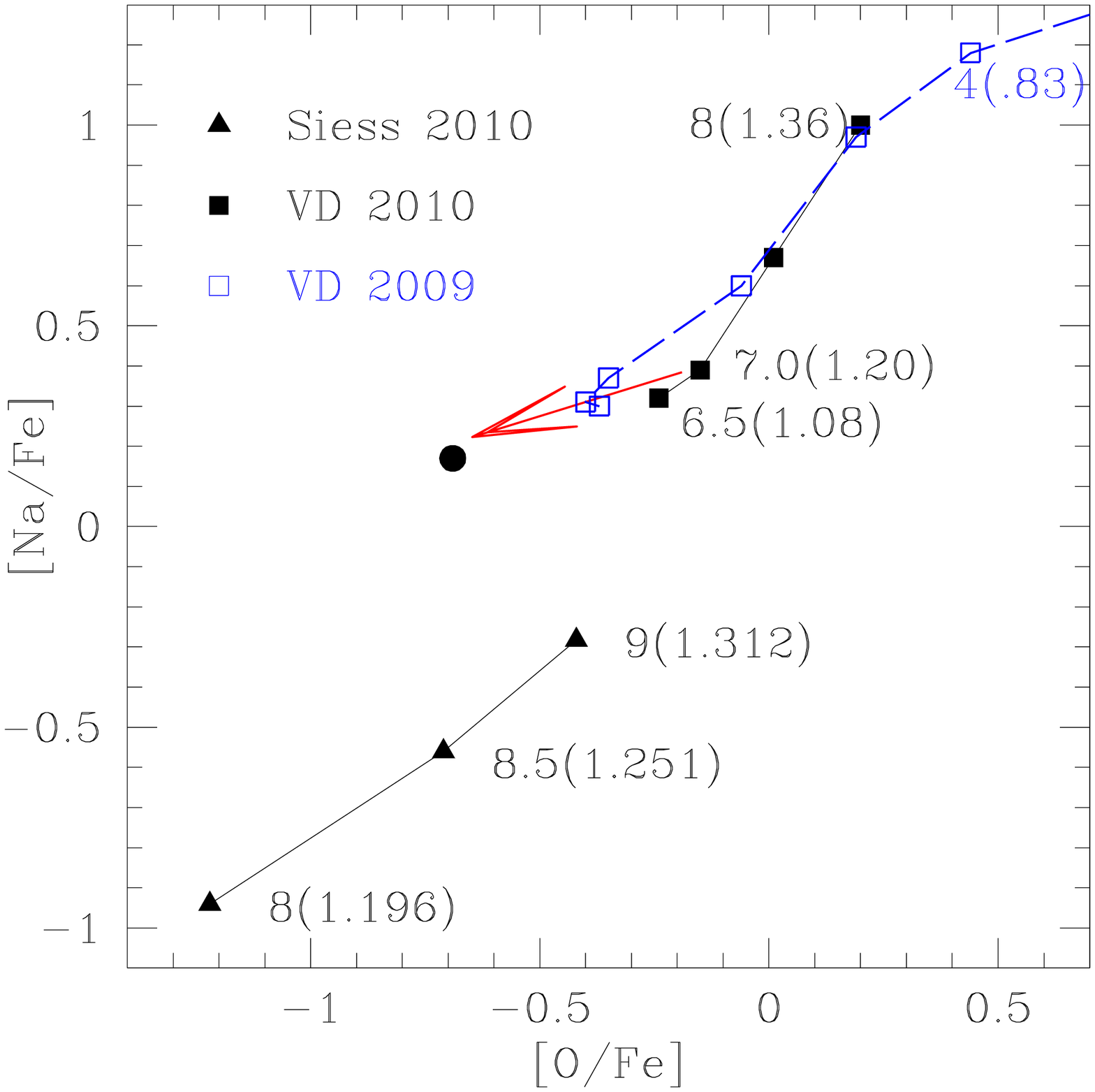}}
\end{minipage}
\caption{The average [O/Fe] and [Na/Fe] content of the
ejecta of the SAGB models by \citet{siess10} (full triangles),
and those calculated in this work (full squares). Open squares denote the
massive AGB yields by \citet{paolo09}. The numbers
next to the points indicate the initial mass of the model, and the core mass (in parenthesis).
The arrow indicates how the yields of the 7M$_{\odot}$ model 
change when the lower mass loss formulation is adopted (see Table 1). }
\label{conf}
\end{figure}

The recent exploration by \citet{doherty} revealed a great homogeneity
among the results obtained with different codes, that confirm the
results concerning the C-burning phase already present in the literature.
These models are limited to the evolutionary phases preceeding the
first thermal pulse, whereas as far as we know the only full investigation
extended to the whole SAGB evolution was given by \citet{siess10}.

Fig.~\ref{conf} shows a comparison between the yields of SAGB models
by \citet{siess10} (indicated as full triangles) and those found in 
the present investigation (open squares), in the [Na/Fe] vs [O/Fe] plane. 
In both cases the points move towards the top-right of the plane
with increasing mass. The yields by \cite{siess10} show lower contents of
oxygen and sodium. The differences in [O/Fe] are mostly due to the
different modelling of mass loss. The treatment by \citet{VW93}, used
by \cite{siess10}, favours much smaller rates, thus leading to a very
advanced nucleosynthesis at the bottom of the outer convective zone.
Conversely, the recipe by \citet{blo} determines much 
higher $\dot{M}$s, so that the mass of the envelope is lost
before a great destruction of oxygen can be achieved. This is confirmed
by the full dot in Fig.\ref{conf}, that indicates the yield 
found in the 7M$_{\odot}$ model when the \citet{VW93} rate is adopted:
the theoretical point is shifted in the range of oxygen abundances 
predicted by \citet{siess10}. The remaining difference is due to the fact that 
our initial mixture is alpha-enhanced with $[\alpha/Fe]=+0.4$,
whereas the mixture used by \citet{siess10} is solar-scaled.
The sodium yields remain different, our [Na/Fe]
being sistematically larger. This can be attributed to the different
nuclear reaction rates adopted, in particular for what concerns the
proton capture reaction by $^{22}$Ne nuclei, for which we adopted
the highest rates from the \citet{hale1} compilations, whereas the
recommended rate was used by \citet{siess10}.

The comparison for what concerns the magnesium and aluminium content
is more tricky, because of the differences in the assumptions made,
that involve the nuclear cross-sections, the initial magnesium
abundances, and the mass loss description. The Al yields by
\citet{siess10} decrease with mass, starting form
[Al/Fe]=0.27 for the M=8M$_{\odot}$ model, down to 
[Al/Fe]=0.18 for M=9M$_{\odot}$. Our yields show the same trend
(see col.8 in Table \ref{sagbmod}), but the values are sistematically
higher, mainly because our alpha-enhanced mixture has a higher
initial content of magnesium. 

This investigation indicates that the oxygen and sodium yields 
in SAGBs vary considerably according to the treatment of mass loss:
when a modest mass loss is adopted, extreme chemistries are
possible, because a very advanced nucleosynthesis can be achieved
at the bottom of the convective envelope. Convective modelling is
also expected to play a role in determining the yields of SAGBs
(see the discussion in \citet{siess10}), although the effects are 
less important than for AGBs, because a more efficient convective modelling
would determine higher temperatures, but also a larger mass loss,
and the two effects tend to compensate for what concerns the
stage of nucleosynthesis achieved.

In the near future, it could be possible to calibrate the mass 
loss rate by achieving constraints from the abundances of stars 
belonging to the bMS in the clusters \ocen and NGC~2808, in the
hypothesis that the bMS stars are directly formed from the ashes of 
SAGBs \citep{pumo2008, annibale}.

\section{Conclusions}
We present the evolution of the main physical and chemical properties
of stars with initial mass in the range 6.5-8M$_{\odot}$, which, according
to our choice for the overshooting from the convective core during the 
H- and He-burning phases, undergo carbon ignition in conditions of partial degeneracy.
Our findings concerning the modalities of carbon burning, the inwards 
propagation of the convective flame that is developed, and the final 
chemistry of the O-Ne core, are in agreement with the results presented
by other research groups.

The same homogeneity cannot be expected for the following TP phase, because
its physical and chemical evolution is determined by the treatment of two
still poorly known physical processes: mass loss and turbulent convection.
We focused our analysis on the chemistry of the ejecta of SAGBs, and in
particular on the oxygen, sodium and aluminum content, that have been
shown to vary among stars in globular clusters.

The models presented here attain extremely large temperatures at the 
bottom of the convective envelope, which, in turn, favour an advanced 
nucleosynthesis, i.e. very small abundances of oxygen and sodium, and a strong 
production of aluminum. 
These findings, which are in fairly good agreement with the results obtained 
by \citet{siess10}, hold in case that mass loss is modelled according to 
\citet{VW93}. Conversely, when the \citet{blo} recipe is used, the 
nucleosynthesis at the bottom of the external mantle is halted by the 
general cooling of the structure favoured by the fast consumption of the 
envelope, so that a less extreme chemistry is achieved. In this latter 
case higher abundances of oxygen and sodium are produced, and, in the overall 
context of a possible pollution in GCs produced by massive AGBs and super-AGBs, 
gas with the most extreme chemistry is expected to be ejected by masses around 
6M$_{\odot}$. 

Helium is produced in great quantities in all cases, with a maximum of 
Y$\sim 0.36$ for the highest masses. In the framework of our models, 
abundances larger than this value for the extreme populations
in the most massive clusters cannot be explained by formation from SAGB ashes.
However, the values derived from observations are still largely uncertain 
\citep{portinari2010}.

If the bMS stars in massive GCs are formed directly from the ejecta of SAGBs, 
future spectroscopic analysis will help understanding the reliability of the 
self-enrichment scenario by massive AGBs, and, eventually, will help to calibrate the
poorly constrained mass loss in the SAGB phase. In the end, this calibration
may also be important to constrain the mass range in which we expect that the
core grows up to begin electron capture on the Ne nuclei. Indirect informations
on the formation of neutron stars by the e-capture supernova channel in single
stars will be eventually achieved \citep{poelarendes}.

\section*{Acknowledgments}
The authors are indebted to the referee, Lionel Siess, for the many
detailed comments and suggestions, that helped improving the quality
of the manuscript.

\end{document}